\documentclass[aps,prx,twocolumn,superscriptaddress,longbibliography]{revtex4-1}

\usepackage{graphicx}
\usepackage{amsmath}
\usepackage{color}
\usepackage[colorlinks=true,linkcolor=blue,urlcolor=blue,citecolor=blue]{hyperref}
\usepackage{hyperref}
\usepackage{braket}
\usepackage{amssymb}
\usepackage{braket}

\begin{document}

\title{Signatures of Many-Body Localization in a Controlled Open Quantum System}

\author{Henrik P. L\"uschen}
\affiliation{Fakult\"at f\"ur Physik, Ludwig-Maximillians-Universit\"at M\"unchen, Schellingstr.\ 4, 80799 Munich, Germany}
\affiliation{Max-Planck-Institut f\"ur Quantenoptik, Hans-Kopfermann-Str.\ 1, 85748 Garching, Germany}

\author{Pranjal Bordia}
\affiliation{Fakult\"at f\"ur Physik, Ludwig-Maximillians-Universit\"at M\"unchen, Schellingstr.\ 4, 80799 Munich, Germany}
\affiliation{Max-Planck-Institut f\"ur Quantenoptik, Hans-Kopfermann-Str.\ 1, 85748 Garching, Germany}

\author{Sean S. Hodgman}
\affiliation{Fakult\"at f\"ur Physik, Ludwig-Maximillians-Universit\"at M\"unchen, Schellingstr.\ 4, 80799 Munich, Germany}
\affiliation{Max-Planck-Institut f\"ur Quantenoptik, Hans-Kopfermann-Str.\ 1, 85748 Garching, Germany}
\affiliation{Research School of Physics and Engineering, Australian National University, Canberra ACT 0200, Australia}

\author{Michael Schreiber}
\affiliation{Fakult\"at f\"ur Physik, Ludwig-Maximillians-Universit\"at M\"unchen, Schellingstr.\ 4, 80799 Munich, Germany}
\affiliation{Max-Planck-Institut f\"ur Quantenoptik, Hans-Kopfermann-Str.\ 1, 85748 Garching, Germany}

\author{Saubhik Sarkar}
\affiliation{Department of Physics and SUPA, University of Strathclyde, Glasgow G4 0NG, Scotland, UK}

\author{Andrew J. Daley}
\affiliation{Department of Physics and SUPA, University of Strathclyde, Glasgow G4 0NG, Scotland, UK}

\author{Mark H. Fischer}
\affiliation{Department of Condensed Matter Physics, Weizmann Institute of Science, Rehovot 7610001, Israel}

\author{Ehud Altman}
\affiliation{Department of Condensed Matter Physics, Weizmann Institute of Science, Rehovot 7610001, Israel}
\affiliation{Department of Physics, University of California, Berkeley, CA 94720}

\author{Immanuel Bloch}
\affiliation{Fakult\"at f\"ur Physik, Ludwig-Maximillians-Universit\"at M\"unchen, Schellingstr.\ 4, 80799 Munich, Germany}
\affiliation{Max-Planck-Institut f\"ur Quantenoptik, Hans-Kopfermann-Str.\ 1, 85748 Garching, Germany}

\author{Ulrich Schneider}
\affiliation{Fakult\"at f\"ur Physik, Ludwig-Maximillians-Universit\"at M\"unchen, Schellingstr.\ 4, 80799 Munich, Germany}
\affiliation{Max-Planck-Institut f\"ur Quantenoptik, Hans-Kopfermann-Str.\ 1, 85748 Garching, Germany}
\affiliation{Cavendish Laboratory, University of Cambridge, Cambridge CB3 0HE, UK}

\date{\today}

\begin{abstract}
In the presence of disorder, an interacting closed quantum system can undergo many-body localization (MBL) and fail to thermalize. However, over long times even weak couplings to any thermal environment will necessarily thermalize the system and erase all signatures of MBL. This presents a challenge for experimental investigations of MBL, since no realistic system can ever be fully closed.
In this work, we experimentally explore the thermalization dynamics of a localized system in the presence of controlled dissipation. Specifically, we find that photon scattering results in a stretched exponential decay of an initial density pattern with a rate that depends linearly on the scattering rate. We find that the resulting susceptibility increases significantly close to the phase transition point. In this regime, which is inaccessible to current numerical studies, we also find a strong dependence on interactions. Our work provides a basis for systematic studies of MBL in open systems and opens a route towards extrapolation of closed system properties from experiments.
\end{abstract}


\maketitle

\section{INTRODUCTION}
\begin{figure*}
	\centering
	\includegraphics[width=175mm]{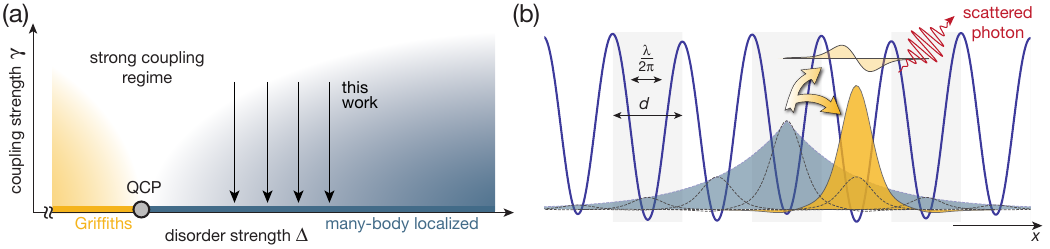}
	\caption{\small{Schematic phase diagram of an open MBL system and illustration of the effects of a single photon scattering event. (a) Coupling a disordered system to a thermal environment destroys MBL on a timescale inversely proportional to the coupling strength $\gamma$. Nonetheless, for sufficiently weak couplings the characteristics of both the MBL (i.e.\ persisting density pattern) and the ergodic Griffiths phase (i.e.\ power law decay of density pattern~\cite{Potter15,Agarval15,Luitz16}) survive sufficiently long to enable their experimental characterization. The shaded area represents the regime of weak dissipation in which the intrinsic dynamics of the phases can be discerned. Due to diverging timescales at the critical point, the respective signatures become increasingly difficult to observe in its proximity. A sharp transition is only expected in the closed system limit ($\gamma = 0$), while finite dissipative couplings are expected to produce a crossover regime between the two characteristic dynamics, similar to the effect of temperature in ground state quantum phase transitions. Black arrows indicate the regime that is considered in this work. (b) Schematic of a scattering event. An atom in an initial localized superposition of Wannier states (blue) becomes localized on the length scale of the scattered photon's wavelength $\lambda/2\pi$, which, in our system, is less than the lattice constant $d$. This dephases superpositions to incoherent mixtures of single Wannier states in the ground band (yellow), as well as producing a small population in higher bands (faint yellow), which can be seen as the result of position measurements with sub lattice site resolution. Band excitations can then lead to atom loss, since in most excited bands atoms are not trapped.}}
	\label{schematic_fig}
\end{figure*}

In a perfectly closed system, many-body localization (MBL) presents a novel paradigm of time evolution, in which quantum correlations can persist locally to arbitrarily long times~\cite{Basko06,Polyakov05,Imbrie14,Oganesyan05,Pal10,Vosk13,Shlyapnikov15,Nandkishore15,Altman15,Bahri15,Serbyn14,Ovadia15,Schreiber15,Bordia15,Smith15}. It thereby provides a robust alternative to conventional thermalizing dynamics. However, experiments are invariably coupled, at least weakly, to their environment, and thus cannot realize a strictly closed system. It is therefore crucial to understand how such dissipative couplings affect the many-body dynamics.
Recently, experiments have observed MBL through the persistence of initially prepared density or spin patterns on intermediate to long timescales~\cite{Schreiber15,Bordia15,Smith15,Choi16,Bordia16}. At even longer times, however, the patterns vanish and the systems become thermal due to residual couplings to the environment. These couplings not only set a timescale for thermalization, but are also expected to broaden the localization transition into a crossover (Fig.~\ref{schematic_fig}(a)), in which the dynamics smoothly interpolate between those of an ergodic and an MBL system. Features of the critical point could then be encoded into a universal dependence of the relaxation curves on the dissipation rate. 
This is similar to the role of temperature in ground state quantum phase transitions, where measuring universal temperature dependencies allows for a characterization of the critical point~\cite{Sachdev_book}.
Systematically varying the strength of the dissipative couplings promises the analogous possibility of extrapolating to closed systems.

Several theoretical works have recently addressed different aspects of MBL systems coupled to an external bath~\cite{BNowak12,Nandkishore14,Gopalakrishnan14,Johri15,Huse15,Nandkishore15,Nandkishore15_2,Banerjee15,Levi15,Fischer15,Znidaric15,Medvedyeva15,Everest16,Nandkishore16}. 
In particular, recent analytical and numerical studies considered the relaxation of almost local integrals of motion associated with MBL under the influence of a weak coupling to a photon bath~\cite{Levi15,Fischer15,Znidaric15,Medvedyeva15,Everest16}. This constitutes a Markovian heat bath at infinite temperature operating mainly through two dissipation channels (Fig.~\ref{schematic_fig}(b)):
(i) effectively measuring the position of the particles and (ii) particle loss. Position measurements affect the system by dephasing coherent superpositions of Wannier states, and therefore show a much stronger effect on systems with finite or longer range quantum coherences as compared to e.g.\ certain glasses that may behave classically already at the scale of the inter-particle distance.

Here, we make use of the exceptional microscopic understanding and control over dissipative processes that is possible with cold atoms in optical lattices~\cite{Mueller12,Daley14} to explore MBL in an open quantum system. Specifically, we investigate how photon scattering affects the dynamics of a density pattern in the many-body localized regime. 
Deep in the localized phase, we find a stretched exponential relaxation of the density pattern in good agreement with theoretical studies~\cite{Levi15,Fischer15,Znidaric15,Medvedyeva15,Everest16}. 
In the experiment, we are furthermore able to study the intriguing regime close to the MBL transition that is not accessible to current numerical studies. There we find that dissipation has an increasingly strong effect that is significantly enhanced by inter-particle interactions.
Our work provides a basis for understanding MBL in realistic (open) experiments and highlights the importance of accounting for the effects of dissipation in order to access the critical point. Furthermore, it demonstrates a versatile tool for systematic studies of open quantum systems, both within and outside of the context of MBL.
\\

\section{Experiment}
We start the experiment by cooling a gas of $130\times10^3$ $^{40}$K atoms in a dipole trap to a temperature of $0.15\,T_F$, where $T_F$ denotes the Fermi temperature. The gas is then loaded into a three dimensional optical lattice consisting of two deep ($\lambda_o = 738.2 \, $nm) orthogonal lattices, which create an array of one dimensional tubes, and a primary ($\lambda_p \approx 532.2\,$nm) lattice with lattice constant $d = \lambda_p/2$ along the tubes. We superimpose the primary lattice with an incommensurate ($\lambda_d \approx 738.2\,$nm) disorder lattice to implement the interacting Aubry-Andr\'e model~\cite{Aubry80,Schreiber15} in the individual tubes. This model describes spinful fermions on a tight-binding lattice with on-site interaction $U$ and nearest-neighbor tunneling amplitude $J \approx h \cdot 500$\,Hz, subject to a quasi-periodic potential $\Delta \cos(2\pi \alpha i + \phi)$ with amplitude $\Delta$. 
Here, $i \in$ $\mathbb{Z}$ numbers the lattice sites, $\alpha=\lambda_p / \lambda_d$ is the disorder periodicity and $\phi$ is the relative phase between the primary and the incommensurable disorder lattice.
In the absence of interactions this model exhibits a localization transition at $\Delta = 2 \, J$~\cite{Aubry80}, and it has been shown to be many-body localized at $U \neq 0$ above a parameter dependent critical disorder strength~\cite{Schreiber15}.

Using a period-two superlattice, we artificially create a charge-density wave state with an initial imbalance $\mathcal{I}=\left ( N_{\rm e} - N_{\rm o} \right ) / \left ( N_{\rm e} + N_{\rm o} \right )$ close to one, where $N_{\rm e}$ ($N_{\rm o}$) denotes the number of atoms on even (odd) sites. 
After the desired evolution time, we extract the remaining imbalance using a superlattice band-mapping technique~\cite{Trotzky12}. 
For ergodic systems, the imbalance decays to zero during the evolution, while a persisting imbalance signals localization. 
Further details of the system, the preparation and the readout sequence can be found in references~\cite{Trotzky12,Schreiber15,Bordia15}.

\begin{figure}
	\centering
	\includegraphics[width=84mm]{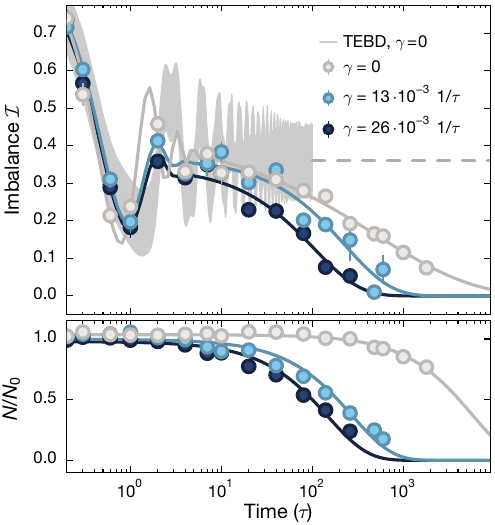}
	\caption{\small{Time evolution of a charge-density wave in the presence of photon scattering.
			Upper panel:  
			An initially prepared one-dimensional charge-density wave evolves in the presence of quasi-periodic disorder of strength $\Delta = 4 \, J$ at $U = 2 \, J$ under the influence of varying scattering rates $\gamma$. Higher scattering rates result in shorter lifetimes of the imbalance. The finite imbalance lifetime at $\gamma = 0$ is due to residual couplings between different 1D tubes~\cite{Bordia15} and off-resonant scattering of lattice photons, which are not included in the TEBD simulation ($\gamma = 0$) indicated by the gray shaded region. The dashed line extrapolates the simulation's mean value.
			Each experimental data point is the average of 6 disorder phase realizations, with errorbars denoting the standard error of the mean.
			Solid lines are stretched exponential fits, used to extract the imbalance decay rates (see Appendix~\ref{fitfunction_appendix}).
			The lower panel shows the corresponding time evolution of the normalized atom number fitted by simple exponentials.}}
	\label{time_traces_fig}
\end{figure}

The atoms are prepared in an equal mixture of the two lowest spin states in the lower ground state manifold with hyperfine quantum number F=9/2. Photon scattering is introduced via a dedicated $\pi$-polarized plane wave laser beam at a detuning of $1.3\,$GHz below the D$_2$ line (see Appendix~\ref{level_appendix}). Starting from the F=9/2 manifold, the absorption and reemission of a photon can leave an atom back in its original state, but may also, with a probability of $\approx \,$33\%, excite it to the upper F=7/2 ground state manifold.
The detuning of the scattering beam is chosen such that the light is essentially resonant for atoms in the F=7/2 manifold, resulting in a quick transfer back to the lower manifold by resonant optical pumping, scattering typically one to five additional photons. 
Such `scattering bursts', which start and end in the lower hyperfine manifold, happen on a timescale that is much shorter than the tunneling time $\tau=\hbar / J$. Consequently, we can consider these bursts as single effective scattering events happening at a total rate of $N \cdot \gamma$, where $N$ denotes the atom number and $\gamma$ the resulting single-particle scattering rate, which sets the effective coupling strength to the bath.
We focus on the weak scattering regime ($\hbar \gamma \ll J$), where atoms can freely time evolve under the closed system Hamiltonian between successive scattering bursts. This is in stark contrast to the strong scattering limit, where a quantum Zeno effect would result in the localization of atoms~\cite{Mueller12,Daley14,Patil15}.
Since we are close to the Paschen-Back regime, optical processes couple only weakly to the magnetic quantum number, such that a scattering burst will leave the spin state of the atom mostly unchanged. Details of the scattering bursts as well as the scattering beam are discussed in the Appendices~\ref{level_appendix},\ref{calc_appendix_1} and~\ref{calc_appendix_2}.
\\

\section{Results}
Fig.~\ref{time_traces_fig} shows sample time traces of imbalance $\mathcal{I}$ and atom number $N$ for various scattering rates $\gamma$ at $U=2\,J$ and moderate disorder $\Delta=4\,J$.
As observed previously~\cite{Schreiber15}, the imbalance settles to a plateau at finite imbalance within a few tunneling times. 
In a perfectly isolated system, this finite imbalance would persist for all times, as is indicated by the numerical simulations.
However, residual couplings between neighboring tubes~\cite{Bordia15}, as well as off-resonant scattering of lattice photons limit the imbalance and atom number lifetimes to $\mathcal{O}(10^3 \tau)$ at the chosen parameters. Note that deeper in the localized phase we have observed significantly longer lifetimes~\cite{Bordia15}.

For finite values of $\gamma$ we observe a faster relaxation of the imbalance and an increased atom loss. This can be understood from a microscopic picture, in which scattering a photon results in the measurement of an atom's position on the length scale of the photon's wavelength $\lambda$ (Fig.~\ref{schematic_fig}(b))~\cite{Sarkar14}.
Because of the relative size of the wavelength and the lattice constant $d$, in our experiment each scattering event can be interpreted as projecting the affected atom onto a single lattice site ($\lambda/(2 \pi d) \approx 0.46$)~\cite{Pichler10}. In this process, the probability for finding the atom on a specific final lattice site after the scattering event is given by the squared wavefunction overlap of the corresponding Wannier state with the atom's original state.
This measurement effectively turns any coherent superposition of Wannier states into an incoherent mixture and can be described as dephasing the coherence terms in the initial single-particle density matrix at a rate $\gamma_{\rm dp} = p_\mathrm{dp} \cdot \gamma$, without altering the occupations~\cite{Fischer15,Levi15,Medvedyeva15}. Here $p_\mathrm{dp}$ gives the probability of a scattering burst resulting in a dephasing event, where an atom remains in the lowest band (see Appendices~\ref{calc_appendix_1},\ref{calc_appendix_2}). Crucially, in the weak scattering limit considered in this work, time evolution under the closed system's Hamiltonian allows atoms to evolve into new coherent superpositions between successive scattering events. Since the new superpositions can be centered around a different lattice site than the original superposition, this effectively re-introduces hopping processes.

In addition, the induced measurement of the atom's position on a length scale $\lambda/2\,\pi$ implies a position measurement also within the lattice site, which can excite population to higher Bloch bands at a rate of $\gamma_{\rm ex} = (1-p_\mathrm{dp}) \cdot \gamma$.
These excitations ultimately result in atom loss, since weak trapping and strong tunnel couplings in higher excited bands allow the atoms to quickly tunnel out of the system.
Note, however, that in our system atoms in the lowest longitudinally excited band remain trapped but are delocalized due to the higher tunneling rate (see Appendix~\ref{loss_appendix}). 
Hence, in the presence of interactions, band excitations can contribute to the imbalance decay through both a complex rearrangement of the ground band wavefunction when an atom is excited, as well as through interactions of ground band atoms with delocalized atoms in higher bands.

\begin{figure}
	\centering
	\includegraphics[width=84mm]{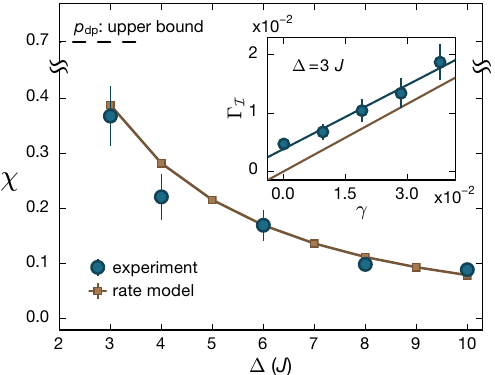}
	\caption{\small{Non-interacting susceptibility vs. disorder strength: 
			Susceptibilities for both the rate model and the experiment. 
			Errorbars indicate the fit uncertainty. 
			The black dashed line indicates the upper bound of $\chi \leq p_\mathrm{dp}$ in the ergodic phase.
			The inset shows measured imbalance decay rates $\Gamma_\mathcal{I}$ as a function of the scattering rate $\gamma$ at $\Delta = 3 \, J$. We observe a linear behavior, the slope of which is directly given by $\chi$.
			We compare the data to the predictions of a rate model~\cite{Fischer15}, indicated by the brown line, which is parallel to the fit through the experimental data. Hence experiment and theory give the same susceptibility. The offset is caused by the constant background decay $\Gamma_\mathrm{bg}$ in the experiment.}}
	\label{suscep_nonint_fig}
\end{figure}

The rates of the dephasing and band excitation processes sum up to the total scattering rate $\gamma$, which is controlled by the intensity of the scattering beam.
In our system, the ratio $\gamma_{\rm dp}/\gamma_{\rm ex} \approx 2.3$ is set by the lattice parameters as well as the wavelength of the scattering beam and is fixed throughout this work. We obtain its value from an ab-initio calculation of a scattering burst, which is discussed in detail in the Appendices~\ref{calc_appendix_1} and~\ref{calc_appendix_2}.

We quantify the imbalance relaxation via fits to a heuristic fit function, which for long times $t$ decays by a stretched exponential of the form $e^{-(\Gamma_{\mathcal{I}}t)^\beta}$~\cite{Levi15,Fischer15}. 
The stretched exponential form arises naturally within a model with a spatial distribution of local relaxation rates~\cite{Fischer15}. 
We find that the global imbalance relaxation rate $\Gamma_{\mathcal{I}}$ increases linearly with $\gamma$, i.e.\ ${(\Gamma_\mathcal{I} - \Gamma_\mathrm{bg}) \propto \gamma}$, in all parameter regimes (inset of Fig.~\ref{suscep_nonint_fig} and Appendix~\ref{scaling_appendix}). This is consistent with an incoherent sum of two independent decay processes, namely the previously studied constant background decay $\Gamma_{\rm bg}$~\cite{Bordia15} and the effects of photon scattering ($\propto \gamma$).

Motivated by the above observation we parameterize the imbalance relaxation rate as
${\Gamma_\mathcal{I}=\chi \cdot \gamma +\Gamma_\mathrm{bg}}$.
For our system the susceptibility to photon scattering $\chi$ depends on the intrinsic system parameters ($U,\Delta$), as well as the ratio of dephasing to excitation processes. 
Intuitively, $1/\chi$ is a measure of the stability of a localized system to the effects of photon scattering and directly relates to the color gradients in Fig.~\ref{schematic_fig}a, with higher values of $\chi$ corresponding to a steeper gradient along $\gamma$. We first analyze the non-interacting case as it is exactly solvable and can be used to calibrate the experiment,
before proceeding to the interacting case.
\\

\subsection{Non-interacting case}
In the absence of interactions, we expect excitations of atoms to higher bands to have no effect on the imbalance, as they occur on even and odd sites with identical rates, and cannot affect the remaining atoms in the non-interacting case.
Fig.~\ref{suscep_nonint_fig} shows the non-interacting susceptibility $\chi$ as a function of disorder strength in the single-particle localized regime ($\Delta > 2 \, J$). 
The susceptibility strongly decreases for increasing disorder strength, which can be understood by considering a single particle localized around a site $i$:
Deep in the localized phase, its time averaged density distribution will be almost identical to that of the Wannier state on site $i$ with almost no weight on neighboring sites. 
In this limit, a photon scattering event has negligible probability of moving the particle away from site $i$, resulting in a vanishing susceptibility. 
At weaker disorder strength, single-particle eigenstates are less localized and have finite overlap with the Wannier states of the neighboring sites. 
Hence, there is now a finite probability of scattering induced hopping transferring the particle to a neighboring site and thereby relaxing the imbalance, giving rise to a finite susceptibility. 
This intuitive idea is also at the heart of a recently proposed rate model~\cite{Fischer15}, which we compare to our data (Fig.~\ref{suscep_nonint_fig}). Since the rate model describes only dephasing events, its scattering rate has been rescaled by $p_\mathrm{dp}$ to take their finite probability into account. 
We find very good agreement between experiment and theory.
This demonstrates that atom losses and excited band populations cannot affect the imbalance in the absence of interactions.

Our observable does not allow us to characterize the susceptibility at disorder strengths below $\Delta \lesssim 3\,J$, since close to the phase transition point the localization length becomes too large and the stationary imbalance of the closed system is already close to zero. 
However, we can derive a simple upper bound for the susceptibility based on the rate equation model:
When the localization length diverges, each dephasing event has equal probability to project the atom onto an even or odd site, thereby canceling its contribution to the imbalance.
In this limit, the imbalance thus decays with the rate $\gamma_{\rm dp}$, giving an upper bound to the susceptibility of $\chi \leq p_\mathrm{dp}$.
\\

\begin{figure}
\centering
\includegraphics[width=84mm]{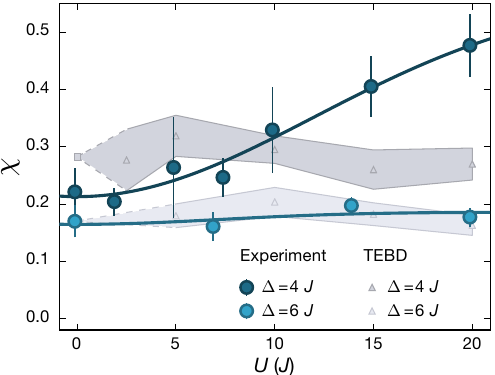}
\caption{\small{ Measured susceptibilities at different disorder strengths versus interactions $U$. At finite interaction strengths, we compare our results to numerical TEBD simulations that do not include particle loss (triangles). The theoretical values at $U=0$ are calculated from the rate model discussed earlier (squares).
We observe a strong interaction dependence of the experimental susceptibilities close to the phase transition ($\Delta = 4 \, J$), but only a weak effect deep in the localized phase ($\Delta = 6 \, J$).
Errorbars indicate the fit uncertainty. The solid lines are guides to the eye and the gray shaded region indicates the statistical uncertainty of the TEBD simulations.}}
\label{suscep_int_fig}
\end{figure}

\subsection{Interacting case}
In the interacting case we expect a higher susceptibility, since now any dephasing event can also affect particles close by. Additionally, atom losses will now perturb the surrounding atoms and thereby further increase the susceptibility~\cite{Fischer15}. On top of these purely dissipative effects, also any delocalized atoms in excited bands can interact with the ground band. Hence, the non-interacting limit $\chi \leq p_\mathrm{dp}$ no longer applies. In analogy to divergent susceptibilities at other phase transitions, one might in fact expect a divergent behavior at the MBL transition. As a consequence, even infinitesimally small couplings would dominate the dynamics close to the critical point, as is indicated in Fig.~\ref{schematic_fig}(a).

Fig.~\ref{suscep_int_fig} shows the measured susceptibilities, as well as the results of TEBD simulations as a function of interaction strength. As in the non-interacting case, the numerical simulation does not implement atom loss, and hence its scattering rate has been rescaled accordingly. In the presence of interactions we do expect this to result in deviations between the experimental and numerical susceptibility, since excitations to higher bands will affect the imbalance.
Deep in the localized phase, at $\Delta = 6\, J$, we observe only a weak effect of interactions consistent with earlier works suggesting that interactions become less important at very strong disorder strengths~\cite{Schreiber15}. In this regime we also find good agreement between theory and experiment, suggesting that atom losses only marginally affect the imbalance.
However, at $\Delta = 4\, J$ we experimentally observe a strongly increasing susceptibility for growing interaction strengths, a trend that we expect to saturate at even larger $U$.  
This is in stark contrast to the TEBD result, which again approaches its non-interacting value at large $U/J$. While the TEBD simulations at $U = 4\,J$ suffer from large truncation errors (see Appendix~\ref{TEBD_appendix}) and hence need to be considered with care, returning to the non-interacting value is the behavior expected for hardcore fermions, due to an exact mapping between the respective Hamiltonians~\cite{Schreiber15}. However, this mapping breaks down when particle numbers are not conserved, suggesting that the difference between experiment and the TEBD simulations are most likely due to the effects of particles being excited to higher bands.
Experimentally disentangling the respective contributions of dephasing and particle excitations is not possible in our setup due to the fixed ratio of $\gamma_\mathrm{dp}/\gamma_\mathrm{ex}$.

An additional challenge is to unravel the effects of pure particle loss from the effects of the trapped but delocalized atoms accumulating in the first excited band. These atoms present an interesting field for future work, since they implement a `small' bath, the properties of which might be strongly influenced by the back-action from the MBL system in the ground band~\cite{Li15,Modak15,Hyatt16}.
\\

\section{Conclusion}
We have realized a controlled open MBL system by introducing dissipation via photon scattering and have found a stretched exponential decay of an initially imprinted charge-density wave in qualitative agreement with recent numerical studies~\cite{Levi15,Fischer15,Znidaric15}.
Systematically varying the scattering rate $\gamma$ enabled us to characterize the robustness of the MBL system via the definition of a susceptibility $\chi$, which we found to be essentially independent of the interaction strength deep in the localized phase.
Furthermore, we were able to experimentally study the interesting regime close to the MBL transition that is not accessible to current numerical studies, and have found an increasing susceptibility upon approaching the critical point. For the non-interacting system we derive an upper bound of $\chi \leq p_\mathrm{dp}$. However, we have found that interactions dramatically increase the system's susceptibility and speculate that they might even cause it to diverge at the MBL transition point, such that even infinitesimally small couplings would dominate the dynamics.

Our study paves the way towards a systematic characterization of the critical point by extrapolating the dynamics at finite coupling to the closed system limit.
A complementary study in the ergodic Griffiths regime, where power-law decays of the imbalance are expected~\cite{Potter15,Agarval15,Luitz16}, would give insight into the delocalized side of the MBL transition.
Furthermore, the applied scheme of implementing open quantum systems via controlled photon scattering is rather general and can straightforwardly be generalized to interesting delocalized states such as e.g.\ superfluids or topological insulators, where controlled dissipation appears to be essential to change the effective Chern number of a state~\cite{Caio15,Alessio15}.

\begin{acknowledgments}
We acknowledge helpful discussions with Anton Buyskikh and Jorge Yago.
We acknowledge financial support by the European Commission (UQUAM, AQuS) and the Nanosystems Initiative Munich (NIM). Work at Strathclyde is supported by the EOARD via AFOSR grant number FA2386-14-1-5003. This research was
supported in part by the National Science Foundation under Grant No. NSF PHY11-25915. MHF acknowledges additional support from the Swiss Society of Friends of the Weizmann Institute of Science.
\end{acknowledgments}

%\cleardoublepage

\appendix

\section{Level scheme and scattering bursts}
\label{level_appendix}

We prepare the $^{40}$K atoms in our system in the lower lying $F=9/2$ hyperfine manifold of their electronic ground state $4$ $^2$S$_{1/2}$, in an equal mixture of $m_F = -9/2$ ($\ket{\downarrow}$) and $m_F = -7/2$ ($\ket{\uparrow}$). We control the interaction strength $U$ between our spins $\ket{\downarrow}$ and $\ket{\uparrow}$ using a \mbox{Feshbach} resonance centered around 202.1\,G~\cite{Regal03}. At these magnetic fields the level structure is close to the Paschen-Back regime, where $m_j$ and $m_i$ become good quantum numbers. This suppresses transitions between different $m_i$ states due to optical transitions to below 10\% (See Appendix~\ref{spin_flip_appendix}). Hence, we can restrict the discussion of optical transitions to the quantum number $m_j$. 

\begin{figure}
	\centering
	\includegraphics[width=84mm]{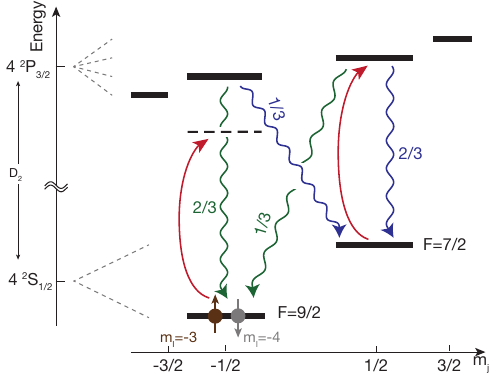}
	\caption{\small{Level scheme of $^{40}K$: 
			Schematics of the atomic hyperfine levels and transitions relevant to the scattering of photons. 
			Levels are labeled using both their quantum number $m_j$, as well as by their respective hyperfine manifold $F$. Here $F$ labels the manifold that the state is adiabatically connected to at low magnetic fields.
			The quantum number $m_i$ is, to good approximation, not coupled to optical transitions as the system is close to the Paschen-Back regime. The two spin states $\ket{m_I=-3}$ and $\ket{-4}$ adiabatically connect to the $\ket{m_F = -7/2}$ and $\ket{-9/2}$ states at low magnetic fields.
			The dedicated scattering light is shown as red arrows, spontaneous emission processes are indicated as wavy lines, along with their branching ratios.}}
	\label{methods_fig}
\end{figure}

A level scheme illustrating all levels and transitions important to the scattering of photons from our dedicated scattering beam is illustrated in Fig.~\ref{methods_fig}. As is indicated by the red arrows, the scattering beam is $\pi$-polarized and its frequency is chosen such that atoms in the $F=9/2$ ground state manifold see a detuning of roughly $1.3\,$GHz to the D$_2$-line, while the upper ground state manifold ($F=7/2$) is coupled resonantly to the excited states.

We define a scattering burst as a series of absorption and reemission processes of photons, where an atom both starts and ends in the lower lying $F=9/2$ manifold. After absorption of an initial photon from the scattering beam, an atom can, via reemission, directly decay back into the lower lying $F=9/2$ manifold, thereby ending the scattering burst, or, with a 33\% probability, decay into the upper $F=7/2$ manifold. Atoms that decayed to the $F=7/2$ manifold experience resonant light and will be excited again. They will therefore quickly scatter multiple photons until, with 33\% probability per scattering, they decay back into the $F=9/2$ manifold, which ends the scattering burst. Hence, a scattering burst typically involves around 1-5 scattered photons.

Since the excitation from the upper $F=7/2$ manifold is resonant, scattering rates from this state are much higher than tunneling rates in the lattice, and hence the total duration of a scattering burst is much shorter than a tunneling time. This means that atoms effectively remain frozen during a scattering burst. Therefore, considering the measurement, or dephasing, effect of a photon scattering, we can treat a full scattering burst as a single effective scattering event. Since the probabilities of exciting an atom into higher bands of the lattice increases with the number of scattered photons in a burst we use an average band excitation probability for the scattering bursts (see Appendix~\ref{calc_appendix_2}).

The total rate of scattering bursts is controlled by the rate of absorbing the first photon from the lower lying $F=9/2$ manifold. On this transition, the detuning and the low light intensities of below $3.6\, \mu$W/cm$^2$ result in scattering rates of only a few scattering bursts per atom per 100 tunneling times.

\section{Single photon band excitation probabilities}
\label{calc_appendix_1}

Scattering photons gives rise to two processes: dephasing of atoms in the ground band and excitation of atoms to higher bands.
Understanding the relative rates of ground band dephasing and excitations associated with the scattering bursts discussed in this work requires a detailed understanding of the processes associated with a single photon. 
We can calculate the single photon rates by calculating the band excitation probabilities of a stimulated absorption followed by a spontaneous emission. 
In this picture, dephasing will be associated with atoms remaining in the ground band. While in highly excited bands atoms will quickly be lost from the trap, the first excited band is an exception, as atoms in this band are trapped (see Appendix~\ref{loss_appendix}).

The calculations are analogous to previous work on heating of atoms in dipole traps~\cite{Gordon1980,Dalibard1985,Gerbier2010,Pichler10,Sarkar14}, and are performed on the combination of three lattices along the three spatial directions for our system parameters. 
The primary lattice along the $x$-axis with $\lambda_p = 532.2$\,nm has a depth of $8 \, E_R^p$. 
The orthogonal lattices ($\lambda_o = 738.2$\,nm) along the $y$- and $z$-axis have a depth of $36 \, E_R^o$. 
Here $E_R^i=h^2/2m\lambda_{i}^2$ is the recoil energy corresponding to the wavelength of the lattice laser $\lambda_{i}$ and atomic mass $m$. 
For this calculation, we will neglect the weak ($<1E_R^d$) disorder lattice, assuming that it only marginally influences the bandstructure.

We calculate the excitation probabilities for atoms starting in a Wannier state of band $(i_x,i_y,i_z)$. Stimulated absorption provides a momentum kick of $\hbar \vec{k}$ along the longitudinal direction, which is the direction of our scattering beam. Here $\vec{k}$ is the momentum of a photon from the dedicated scattering beam. 
Afterwards we act with another momentum kick of $\hbar k$, along an arbitrary direction to model spontaneous emission.
The results of extracting the final excitation probabilities from band $(i_x,i_y,i_z)$ into band $(j_x,j_y,j_z)$, averaged over all emission directions, are shown in Table~\ref{single_phot_table}.

Note that the excitation probabilities into the excited bands of the orthogonal lattices are equal due to symmetry and much lower than the excitation probabilities in the $x$-direction. This is due to the orthogonal lattices being deeper, and the momentum kicks from absorption of photons being along the x-direction due to the direction of travel of our dedicated scattering beam. 

All estimations for the scattering bursts are based on the results of this calculation. Note that we are making small simplifications by using a Wannier state as the starting state, instead of the actual localized wavefunctions, which are a superposition of few Wannier states. Thereby we neglect the potential small effects from coherences. Also, we neglect return processes from `higher' bands that are not explicitly considered, since atoms in these bands are not trapped.

\begin{table}
	\begin{ruledtabular}
		\begin{tabular}{ | c ||c | c | c | c | c | }
			\hline 
			& \parbox[t]{1.4cm}{($j_x,j_y,j_z$)\\=(0,0,0)} & (1,0,0) & (0,1,0) & (0,0,1) & (2,0,0) \\
			\hline
			\parbox[t]{1.4cm}{($i_x,i_y,i_z$)\\=(0,0,0)} & 0.823 & 0.103 & 0.023 & 0.023 & 0.016\\
			\hline
			(1,0,0) & 0.103 & 0.582 & 0.003 & 0.003 & 0.229\\
			\hline
			(0,1,0) & 0.023 & 0.003 & 0.772 & 0.000 & 0.000\\
			\hline
			(0,0,1) & 0.023 & 0.003 & 0.000 & 0.772 & 0.000\\
			\hline
			(2,0,0) & 0.016 & 0.229 & 0.000 & 0.000 & 0.406\\
			\hline
			higher & 0.011 & 0.080 & 0.201 & 0.201 & 0.348 \\ \hline
		\end{tabular}
	\end{ruledtabular}
	\caption{\label{single_phot_table}Single photon excitation probabilities from the ($j_x,j_y,j_z$)-th to the ($i_x,i_y,i_z$)-th band. The index (0,0,0) refers to the ground band, higher indices to the $i$-th ($j$-th) excited band along the given spacial direction.}
\end{table}

\section{Band excitation probabilities of scattering bursts}
\label{calc_appendix_2}
\begin{figure}
	\centering
	\includegraphics[width=84mm]{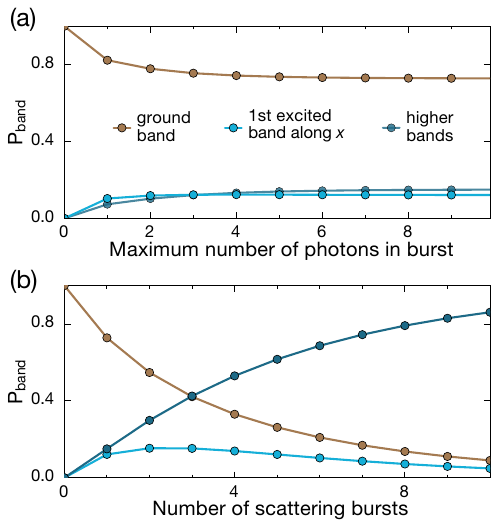}
	\caption{\small{Band excitation probabilities: (a) As a function of maximum possible number of individual photons considered in a scattering burst. Plotted is the integrated band excitation probability for bursts with up to $n$ photons. This integral quickly converges to the average band-excitation probability of a scattering burst. (b) Excitation probability as a function of the number of scattering bursts. The probability of staying in the ground band steadily decreases, while the probability of higher bands, which get lost, increases. At intermediate numbers of scattering bursts a finite population of atoms builds up in the first longitudinally excited band, which is trapped. Time traces used to extract the imbalance decay rate in this work usually contain up to 10 scattering bursts per particle.}}
	\label{scattering_event_data_fig}
\end{figure}

Based on the calculations done for a single scattered photon, we can estimate the average band excitation probability of a scattering burst. 
A scattering burst (see Appendix~\ref{level_appendix}) consists of a first photon absorbed from the $F=9/2$ manifold, followed by a small number $n \in [0,1,2,3,...]$ of photons scattered from the $F=7/2$ transition, before returning to the $F=9/2$ manifold. 
The band excitation probabilities of such a burst will depend heavily on the number of individual photons involved.

We calculate the average band excitation probability of a scattering burst by averaging over all possible realizations of a burst, characterized by the number of cycled photons $n$. 
Specifically, we sum over the band excitation probabilities after scattering $n$ photons (which steadily increases), weighted by the probability of scattering $n$ photons in a burst, which is given by the geometric series $P(n) = (1/3) \cdot (2/3)^{n-1}$ (which quickly converges to zero). 
This sum is plotted in Fig.~\ref{scattering_event_data_fig}(a) as a function of the maximum number of photons considered in a burst.

We observe a quickly converging behavior of all populations after only a few photons.
The limiting ($n \rightarrow \infty $) values give the average excitation probabilities of a scattering burst.

\section{Atom loss mechanism and finite excited band population}
\label{loss_appendix}
In this work, we distinguish between two effects of a photon scattering burst. 
While dephasing is associated with events where atoms stay in the ground band, atom loss occurs due to particles being excited to higher bands and tunneling out of the system. 
Since atoms are mainly excited into higher bands of the longitudinal ($x$) lattice, we expect tunneling along this direction to constitute the main loss mechanism.

\begin{figure}
	\centering
	\includegraphics[width=84mm]{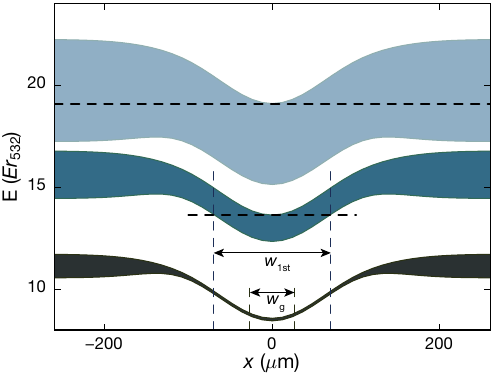}
	\caption{\small{Spatial band structure: Spatial band structure of the lowest three bands of the  longitudinal ($x$) lattice along the $x$-direction at $y,z$=0. The ground band is illustrated in black, the first longitudinally excited band in dark blue and the second longitudinally excited band in light blue. The structure emerges due to the Gaussian shape of the dipole and lattice beams. The red line illustrates the pure trapping potential. Dashed horizontal lines illustrate that atoms in the 2nd excited band can tunnel out of the system, but atoms in the first excited band remain trapped. The cloud size in the ground band is indicated by $w_g$, which corresponds to the $1/e^2$ width of a Gaussian fit. For the 1st excited band, the indicated size $w_{1st}$ is derived from the band structure calculation, which agrees well with the result of an in-situ measurement.
		}}
		\label{bandstructure_fig}
\end{figure}

Fig.~\ref{bandstructure_fig} illustrates the energies of the ground, 1st and 2nd excited band of the longitudinal lattice as a function of real space position.
The Gaussian shape trapping potential stems from the Gaussian beam shapes of the dipole trap.
The bandwidth of the bands increases away from the trap center, because the orthogonal lattice beams also have a Gaussian shape. At a distance of $200\,\mu m$ from the trap center, these beams have zero intensity and hence the atoms only experience the $x$-lattice. 
Vertical dashed lines mark the width of the atom cloud in the ground band, based on an in-situ measurement of the cloud.
Indicated is the full $1/e^2$ width of a Gaussian fit, which corresponds to approximately 200 lattice sites.
Since the ground band is localized via disorder and photon assisted hopping gives only slow, diffusive spreading, we expect the cloud size to remain essentially constant during the dynamics.

In order to enable tunneling out of the system, a band needs to be i) delocalized ($2 \, J_{\rm band} \leq \Delta_{\rm band}$), where $\Delta_\mathrm{band}$ is the disorder strength felt by atoms in the respective band, and ii) untrapped ($4 \, J_{\rm band} > V_{\rm trap}$, where $V_\mathrm{trap}$ is the trap depth).
Due to higher tunnel couplings, the first criterion is true for all longitudinally excited bands.
A graphical visualization of the second condition is illustrated for the 1st and 2nd longitudinally excited band: A horizontal line from the upper band edge must not cross the lower band edge. 
This criterion is fulfilled for the 2nd longitudinally and higher excited bands and hence atoms in these bands can be lost from the system.

In the case of the 1st excited band, however, the line crosses the lower band edge, marking a finite size that atoms in the second band will expand to.
We have checked these predictions by i) measuring the size that the 1st excited band expands to by deliberately loading atoms into the 1st excited band using the superlattice, letting them time evolve and imaging the cloud in-situ, as well as ii) directly measuring the lifetime of the 1st excited band. We obtained good agreement with the predicted size and found a lifetime similar to the lifetime of the ground band in the absence of photon scattering.

Furthermore, calculations of the bandstructure along the orthogonal $(y,z)$ direction shows that, due to the deeper lattices, both the 1st and 2nd excited band are trapped. While photon scattering only excites a few atoms into these bands, they might nonetheless become relevant, since at the spatial edges of the system the first excited bands along $y$ and $z$ are resonant with the first excited band along $x$, enabling transfer of atoms between the bands. We have experimentally checked this by preparing atoms in the first excited band and found that atoms indeed distribute between the 1st excited bands in all three directions.

The 1st excited bands being trapped will result in a finite population in these bands building up.
Fig.~\ref{scattering_event_data_fig}(b) shows the band populations versus the number of scattering bursts. 
The values are calculated using rate equations based on the average excitation probabilities of a scattering burst.
The total population in the 1st excited band quickly builds up to about $15\%$ of the initial atoms before slowly decaying. 
Due to the decay of the ground band population, the 1st excited band population quickly reaches a significant portion of the ground band population, namely approximately 30\% after 4 scattering bursts.

\section{Effects of the finite excited band populations}

Trapped atoms in the excited bands can affect the imbalance in the ground band in multiple ways. The most direct way of influence is due to the imaging procedure not being able to distinguish fully between higher and ground band atoms, which directly affects the measured imbalance.
However, we believe this effect to be small, since the atoms distribute among bands in all directions such that the individual populations are too small and vanish in the noise. Furthermore, we find good agreement with theory in the non-interacting case.

In the presence of interactions one can envisage another possible channel of influence, as atoms in the higher bands, which are delocalized, can act as a bath for atoms in the ground band. While the coupling to this bath should be rather weak due to the bigger spatial size of the 1st longitudinally excited band, it might be significant in certain regimes.
While models describing such a two band behavior have been studied theoretically~\cite{Li15,Modak15,Hyatt16}, those studies have been limited to very small system sizes. A detailed study of the effects of higher band population on the ground band would constitute a particularly interesting future direction for this work.

\section{Calibration of the scattering rates}

In the experiment, we vary the amount of scattering light by controlling the intensity of the scattering beam via an acousto-optic modulator and stabilize the total power using a calibrated photodiode. 
We calibrate the photodiode via the intensity profile of the scattering beam by imaging it at the position of the atoms and comparing it to an in-situ image of the atomic cloud. 
From these images we can obtain the average intensity $I$ at the position of the atoms.
Finally, the scattering rate can be calculated as

\begin{equation}
\gamma = \frac{3 \pi c^2}{2 \hbar \omega_{D_2}^3} \left ( \frac{\Gamma_{D_2}}{ \delta_{\rm{sc}} } \right )^2 I .
\label{scatt_rate}
\end{equation}

\noindent

Here $\omega_{D_2}$ and $\Gamma_{D_2}$ denote the transition frequency and the decay rate of the D$_2$ line, respectively. The detuning $\delta_{\rm sc}$ refers to the detuning seen by atoms in the lower $F=9/2$ hyperfine manifold of the ground state, since the absorption from this state controls the rate of scattering bursts.
Due to the detuning being $\delta_{\rm sc} \approx 1.3\,$GHz, we can neglect the effects of the D$_1$ line, which is much further away, and assume that we do not resolve the hyperfine levels of the excited state, allowing us to use this simple formula.

\subsection{Estimating the relative dephasing rate}

Comparing the experimental data to theory, which only includes the effects of dephasing, requires an estimation of the fraction of scattering bursts resulting only in dephasing $\gamma_{\rm dp}/\gamma = p_\mathrm{dp}$. 
Ignoring back-transfer processes from the 1st excited to the ground band (which would change the ground band population by $\sim 1\%$), this is equal to the probability of staying in the ground band during an average scattering burst, which was calculated earlier. 
This gives a relative dephasing rate of $\gamma_{\rm dp}/\gamma \approx 70$\%.

\subsection{Estimating the relative loss rate}

\begin{figure}
	\centering
	\includegraphics[width=84mm]{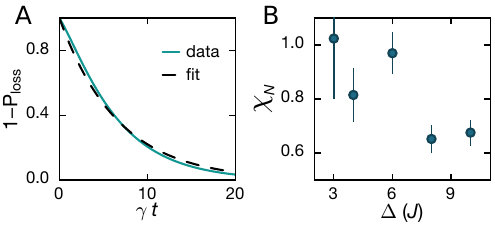}
	\caption{\small{Atom number loss: 
			(a) Fraction of atoms remaining in the ground or 1st excited bands vs time. Dashed line shows an exponential fit to extract the timescale. (b) Non-interacting atom number susceptibilities. Errorbars show the uncertainty of the extracted value, including both the fitting procedure and the error in the initial data.}}
	\label{atom_loss_fig}
\end{figure}

In order to check our calibration of atom loss and the calculations on band excitations we estimate the expected loss rate and compare it to the experimentally measured atom number decay.
Since the 1st excited band along $x$ and the higher bands along the orthogonal directions are trapped, the loss rate should be equal to the rate at which atoms are excited to the 2nd excited band along $x$. 

By summing the exact probabilities of an atom being excited to the 2nd excited band after $n$ scattering bursts (Fig.~\ref{scattering_event_data_fig}(b)), weighted by the probability of scattering $n$ photons in time $t$, which is given by a Poisson distribution

\begin{equation}
P(n,t) = \frac{\left(\gamma t \right)^n e^{- \gamma t}}{n!},
\end{equation}

\noindent
we can calculate the probability of staying in the system (the probability of staying in the ground or 1st excited bands) until time $t$.
While this is not strictly an exponential decay, it can be approximated as such, allowing us to extract an effective loss rate $\gamma_{el} \approx 0.175 \gamma$ (Fig.~\ref{atom_loss_fig}(a)).

Using this rate, the linear relationship between atom loss rate and scattering rate plotted in Fig.~\ref{soms_suscep_def_fig}(b), allows us to define an atom number susceptibility
\begin{equation}
\chi_{_{N}} = d \Gamma_{N} / d \gamma_{el} .
\end{equation}
\noindent
Fig.~\ref{atom_loss_fig}(b) shows the atom number susceptibility for the non-interacting case for various $\Delta$.
We observe a noisy behavior consistent with no trend along $\Delta$. 
Perfect agreement with our model would be indicated by $\chi_{_{N}} =1$. 
We observe values slightly below one, indicating that our model describes excitation processes reasonably well.

One possible explanation for the observed minor differences is the experimental extraction of the atom number lifetime. Since the time traces were only taken up to times where the imbalance reaches zero, the atom numbers have often not fully decayed yet, rendering the exponential fit unreliable.

\section{Spin flip probabilities}
\label{spin_flip_appendix}
At the magnetic fields of around $ 200 \, G$ used in our experiment, the $^2P_{3/2}$ excited state manifold is deep in the Paschen-Back regime. However, the $^2S_{1/2}$ ground state manifold still has a weak coupling between the nuclear and electronic spins, causing a finite probability of changing $m_i$ by scattering a photon.

We can calculate the probability of a spin flip by including nuclear spin in our calculation of scattering rates and branching ratios. We find probabilities of 4\% for $\ket{m_F = -9/2}$ and 10\% for $\ket{m_F = -7/2}$.
Note that most of the spin flips will simply convert atoms between $\ket{m_F = -9/2}$ and $\ket{m_F = -7/2}$. However also $\ket{m_F = -5/2}$ states can be created, which would have a different interaction strength. But given the minimal excitation probabilities, we expect any effects due to these additional spin states to be negligible.

\section{Decay rate scaling with scattering rate}
\label{scaling_appendix}
\begin{figure}
	\centering
	\includegraphics[width=84mm]{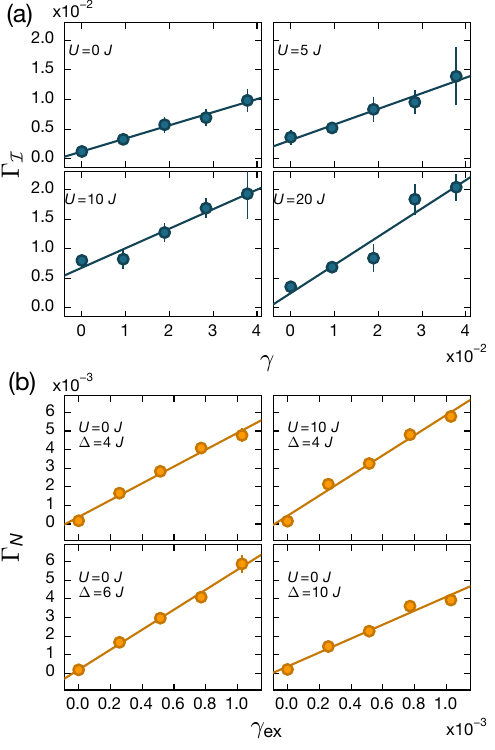}
	\caption{\small{Imbalance and atom number decay rate as a function of the scattering rate: (a) Imbalance decay rate for various interaction strengths at $\Delta=4\,J$. Our data is consistent with a linear behavior for all parameter values. The imbalance decay rates additionally show a background decay rate at $\gamma_\mathrm{dp} = 0$, which heavily depends on both the interaction and the disorder strength. (b) Atom number decay rates as a function of the band excitation rate. Again we find a linear scaling. }}
	\label{soms_suscep_def_fig}
\end{figure}

As discussed in the main paper, the imbalance decay rate shows a linear behavior with the scattering rate. Fig.~\ref{soms_suscep_def_fig}(a) shows further exemplary data for various interaction strengths at $\Delta = 4 \, J$. The susceptibilities are extracted via a linear fit to this data. The errorbars plotted in Figs.~\ref{suscep_nonint_fig},\ref{suscep_int_fig} are calculated as the maximum of i) the square root of the covariance error of the fit and ii) the results of linear fits through the imbalance decay rates plus or minus their respective errorbars. Fig.~\ref{soms_suscep_def_fig}(b) shows experimental data for the atom number decay rate as a function of the calculated band excitation rate $\gamma_\mathrm{ex} = (1-p_\mathrm{dp}) \cdot \gamma$. We observe the expected linear trend, but with a slope  of $\approx$\,0.5, smaller than one. This indicates that not all atoms excited to higher bands are lost. Indeed we find that atoms in the first excited band of the longitudinal lattice remain trapped (see Appendix~\ref{loss_appendix}).

\section{Fitfunction for the Imbalance}
\label{fitfunction_appendix}
The decay of the imbalance is fitted by an initial oscillation that decays exponentially (with amplitude $A$, frequency $\omega$, lifetime $\tau$) to an offset value $o$, which corresponds to the stationary imbalance of the closed system. This function is multiplied by a stretched exponential decay to zero~\cite{Bordia15}, that models the long term decay studied here.

\begin{equation}
\mathcal{I}(t) = \left ( A e^{-t/\tau} \text{cos} \left( \omega t + \phi \right ) + o \right ) e^{-\left ( \Gamma_{\mathcal{I}}t \right ) ^{\beta}}
\end{equation}
Here $\Gamma_\mathcal{I}$ gives the imbalance decay rate and $\beta$ the stretching exponent. We find that a stretched exponential fit describes our data much better than a simple exponential decay. We do not perform a systematical analysis of the stretching exponents $\beta$, since their values depend heavily on the behavior after long times, where the systems response is heavily affected by the effects of trapped atom's in higher bands, a strongly reduced density and the creation of additional spin states. Fit values for $\beta$ scatter between typical values of $0.5 \leq \beta \leq 1$ and show large errorbars.

\section{TEBD simulation of the time evolution}
\label{TEBD_appendix}

Our system can be described by the interacting Aubry-Andr\'e model for spinful Fermions,
\begin{equation}
\hat{H} = -J\sum_{i,s} (\hat{c}^{\dag}_{i, s} \hat{c}^{\phantom{\dag}}_{i+1, s} + \textrm{h.c.}) + \sum_{i,s} V_i \hat{n}_{i,s} + U\sum_{i} \hat{n}_{i,\uparrow}\hat{n}_{i,\downarrow},
\label{eq:HIntAnderson}
\end{equation}
where $\hat{c}^{\dag}_{i,s}$ ($\hat{c}_{i,s}$) creates (annihilates) a particle at site $i$ with spin $s$ and $\hat{n}_{i,s}=\hat{c}^{\dag}_{i,s}\hat{c}^{\phantom{\dag}}_{i,s}$. The disorder potential for the Aubry-Andr\'e model is given by $V_i = \Delta \cos(2\pi \alpha i + \phi)$ with $\alpha$ the ratio of the lattice periodicities and $\phi$ is a random phase.

To simulate the time evolution of the open system, we introduce the density matrix, for which the time evolution is given by the Lindblad equation 
\begin{equation}
\dot{\rho} = -i[H, \rho] + \gamma \sum_{\mu}\Big(L^{\phantom{\dag}}_{\mu}\rho L_{\mu}^\dag - \frac12\{L_{\mu}^\dag L_{\mu}^{\phantom{\dag}}, \rho\}\Big).
\label{eq:lindblad}
\end{equation}
Here, the first term describes the unitary time evolution and the second term the coupling of the system to the environment. The jump operators $L_\mu$ denote the system operators directly coupled to the bath, which in our case are given by local density measurements, i.e., $L_{\mu}=\hat{n}_{i,s}$.  
We then simulate the time evolution of the quantum Lindblad equation~\eqref{eq:lindblad} for system size $S=20$ and 30 disorder realizations using the time-evolving block decimation (TEBD) scheme for matrix product operators~\cite{Zwolak04}.

We note that the results of the numerical calculations, in particular close to the transition and for intermediate interactions, should be treated with some care. Since the local Hilbert-space dimension of the density matrix is $d = 16$, we could not increase the matrix product state bond dimension to more than $100$. For such bond dimension, the truncation error grows rapidly to $\approx 0.1$ per bond before decreasing again. The actual error, however, can not be deduced from this truncation error. The error bars in Fig.~\ref{suscep_int_fig} are the statistical errors from averaging over the random phases $\phi$.

%Bibliography
%

\bibliography{MBL_scattering}
\end{document}